\documentclass[aps,superscriptaddress,preprintnumbers,pra,amsmath,twocolumn,amssymb]{revtex4}%
\usepackage[bookmarks=true, pdfpagemode = None, pdfstartview = FitH,colorlinks = true, citecolor = black, linkcolor = black, urlcolor=black]{hyperref}
\usepackage{graphicx}
\newcommand{\QMA}{{\sf QMA}}
\newcommand{\NP}{{\sf{NP}}}

\def\LH{{\sc local Hamiltonian}}
\def\2LH{{\sc $2$-local Hamiltonian}}
\def\5LH{{\sc $5$-local Hamiltonian}}
\def\AH{{\sc $2$-local ZX Hamiltonian}}
\def\SH{{\sc $2$-local ZZXX Hamiltonian}}
\def\RH{{\sc $2$-local real Hamiltonian}}
\def\RFLH{{\sc $5$-local real Hamiltonian}}
\def\clock{{\mathrm{clock}}}
\def\YES{{\sc Yes}}

\newcommand{\bra}[1]{\langle#1|}
\newcommand{\ket}[1]{|#1\rangle}
\newcommand{\ketbra}[2]{|#1\rangle\langle#2|}
\newcommand{\bydef}{\stackrel{\mathrm{def}}{=}}
\newcommand{\PXi}{\sigma^x_i}
\newcommand{\PYi}{\sigma^y_i}
\newcommand{\PZi}{\sigma^z_i}
\newcommand{\Xj}{\sigma^x_j}
\newcommand{\Yj}{\sigma^y_j}
\newcommand{\Zj}{\sigma^z_j}
\newcommand{\Xk}{\sigma^x_k}

\newcommand{\Zk}{\sigma^z_k}
\newcommand{\X}{\sigma^x}

\newcommand{\Z}{\sigma^z}

\begin{document}

\title{Realizable Hamiltonians for universal adiabatic quantum computers}

\author{Jacob D. Biamonte}\email{jacob.biamonte@comlab.ox.ac.uk}\affiliation{Oxford University Computing Laboratory, \\Wolfson Building, Parks Road, Oxford, OX1 3QD, United Kingdom.}
\author{Peter J. Love}\email{plove@haverford.edu}
\affiliation{Department of Physics, 370 Lancaster Ave., Haverford College, Haverford, PA USA 19041.}

\begin{abstract}
It has been established that local lattice spin Hamiltonians can be used for universal adiabatic quantum computation.  However, the 2-local model Hamiltonians used in these proofs are general and hence do not limit the types of interactions required between spins.  To address this concern, the present paper provides two simple model Hamiltonians that are of practical interest to experimentalists working towards the realization of a universal adiabatic quantum computer.  The model Hamiltonians presented are the simplest known quantum-Merlin-Arthur-complete (\QMA-complete) 2-local Hamiltonians.  The 2-local Ising model with 1-local transverse field which has been realized using an array of technologies, is perhaps the simplest quantum spin model but is unlikely to be universal for adiabatic quantum computation.  We demonstrate that this model can be rendered universal and \QMA{}-complete by adding a tunable $2$-local transverse $\sigma^x\sigma^x$ coupling. We also show the universality and \QMA{}-completeness of spin models with
only 1-local $\Z$ and $\X$ fields and 2-local $\sigma^z\sigma^x$ interactions.

\end{abstract}

\maketitle


What are the minimal physical resources required for universal quantum computation? This question is of interest in understanding the connections between physical and computational complexity, and for any practical implementation of quantum computation.  In 1982, Barahona~\cite{Bar82} showed that finding the ground state of the random field Ising model is \NP{}-hard.  Such observations fostered approaches to solving problems based on classical~\cite{KGV83} and later quantum annealing~\cite{BBRA99}. The idea of using the ground state properties of a quantum system for computation found its full expression in the adiabatic model of quantum computation~\cite{FGGS00}.  This model works by evolving a system from the accessible ground state of an initial Hamiltonian $H_\text{i}$ to the ground state of a final Hamiltonian $H_\text{f}$, which encodes a problem's solution.  The evolution takes place over parameters $s\in[0,1]$ as $H(s) = (1-s) H_\text{i} + s H_\text{f}$, where $s$ changes slowly enough that transitions out of the ground state are suppressed~\cite{AR06}. The simplest adiabatic algorithms can be realized by adding non-commuting transverse field terms to the Ising Hamiltonian: $\sum_ih_i\PZi+\sum_i\Delta_i\PXi+\sum_{i,j}J_{ij}\PZi\Zj$, (c.f.~\cite{2006cond.mat..8253H}).  However, it is unlikely that the Ising model with transverse field can be used to construct a universal adiabatic quantum computer~\cite{BDOT06}.

What then are the simplest Hamiltonians that allow universal adiabatic quantum computation?  For this we turn to the complexity class quantum-Merlin-Arthur (\QMA{}), the quantum analog of \NP{}, and consider the \QMA{}-complete problem k-\LH{}~\cite{KSV02}.  One solves k-\LH{} by determining if there exists an eigenstate with energy above a given value or below another---with a promise that one of these situations is the case---when the system has at most k-local interactions.  A \YES{} instance is shown by providing a witness eigenstate with energy below the lowest promised value.

The problem 5-\LH{} was shown to be \QMA{}-complete by Kitaev~\cite{KSV02}.  To accomplish this, Kitaev modified the autonomous quantum computer proposed by Feynman~\cite{Fey82}.  This modification later inspired a proof of the polynomial equivalence between quantum circuits and adiabatic evolutions by Aharonov \emph{et al.}~\cite{AvDK+05} (see also~\cite{Sui05,MLM06}).  Kempe, Kitaev and Regev subsequently proved \QMA{}-completeness of $2$-\LH{}~\cite{KKR06}.  Oliveira and Terhal then showed that universality remains even when the $2$-local Hamiltonians act on particles in a subgraph of the {\sc 2D} square lattice~\cite{OT06}. Any \QMA{}-complete Hamiltonian may realize universal adiabatic quantum computation, and so these results are also of interest for the implementation of quantum computation.

Since $1$-\LH{} is efficiently solvable, an open question is to determine which combinations of $2$-local interactions allow one to build \QMA{}-complete Hamiltonians.  Furthermore, the problem of finding the minimum set of interactions required to build a universal adiabatic quantum computer is of practical, as well as theoretical, interest: every type of $2$-local interaction requires a separate type of physical interaction.  To address this question we prove the following theorems:\\

\noindent \textbf{Theorem 1.} \emph{The problem \SH{} is \QMA{}-complete, with the ZZXX Hamiltonian given as: }
\begin{eqnarray}\label{eqn:zzxx}
H_{\text{ZZXX}}&=&\sum_{i}h_i\PZi+\sum_{i}\Delta_i\PXi+\\\nonumber
&+&\sum_{i,j}J_{ij}\PZi\Zj+\sum_{i,j}K_{ij}\PXi\Xj.
\end{eqnarray}

\noindent \textbf{Theorem 2.} \emph{The problem \AH{} is \QMA{}-complete, with the ZX Hamiltonian given as}
\begin{eqnarray}\label{eqn:zx}
{H}_{\text{ZX}}&=&\sum_{i}h_i\PZi+\sum_i\Delta_i\PXi+\\\nonumber
&+&\sum_{i<j}J_{ij}\PZi\Xj+\sum_{i<j}K_{ij}\PXi\Zj.
\end{eqnarray}

\paragraph{Structure} In the present paper we briefly review the standard circuit to adiabatic construction to show that $2$-\LH{} is \QMA{}-complete when restricted to real-valued Hamiltonians. We then show how to approximate the ground states of such $2$-local real Hamiltonians by the ZX and ZZXX Hamiltonians.  We conclude this work by providing references confirming our claim that the Hamiltonians in Eq.~\eqref{eqn:zzxx} and~\eqref{eqn:zx} are highly relevant to experimentalists attempting to build a universal adiabatic quantum computer.

\section{The Problem}

The translation from quantum circuits to adiabatic evolutions began when Kitaev~\cite{KSV02} replaced the time-dependence of gate model quantum algorithms with spatial degrees of freedom using the non-degenerate ground state of a positive semidefinite Hamiltonian:
\begin{eqnarray}\label{eqn:htot}
&&0=H\ket{\psi_{\text{hist}}} = \\
&&(H_{\text{in}}+ H_{\clock}+ H_{\text{clockinit}}+ H_{\text{prop}})\ket{\psi_{\text{hist}}}.\nonumber
\end{eqnarray}
To describe this, let $T$ be the number of gates in the quantum circuit with gate sequence $U_T\cdots U_2U_1$ and let $n$ be the number of logical qubits acted on by the circuit.  Denote the circuit's classical input by $\ket{x}$ and its output by $\ket{\psi_{\text{out}}}$.  The {\em history state} representing the circuit's entire time evolution is:
\begin{eqnarray}
\ket{\psi_{\text{hist}}}&=& \frac{1}{\sqrt{T+1}}\biggl[\ket{x}\otimes\ket{0}^{\otimes T}+U_1\ket{x}\otimes\ket{1}\ket{0}^{\otimes T-1}\nonumber\\
&+&U_2U_1\ket{x}\otimes\ket{11}\ket{0}^{\otimes T-2}\nonumber\\
&+&\ldots\\
&+&U_T\cdots
U_2U_1\ket{x}\otimes\ket{1}^{\otimes T}\biggr],\nonumber
\end{eqnarray}
where we have indexed distinct time steps by a $T$ qubit unary clock.  In the following, tensor product symbols separate operators acting on logical qubits (left) and clock qubits (right).

$H_{\text{in}}$ acts on all $n$ logical qubits and the first clock qubit.  By annihilating time-zero clock states coupled with classical input $x$, $H_{\text{in}}$ ensures that valid input state ($\ket{x}\otimes\ket{0...0}$) is in the low energy eigenspace:
\begin{eqnarray}\label{eqn:Hin}
H_{\text{in}}&=&\sum_{i=1}^n (\openone-\ket{x_i}\bra{x_i})\otimes \ket{0}\bra{0}_1\\
&+&\left(\frac{1}{4}\right)\sum_{i=1}^n(\openone -(-1)^{x_i}\PZi)\otimes(\openone+\Z_1).\nonumber
\end{eqnarray}

$H_{\clock}$ is an operator on clock qubits ensuring that valid unary clock states $\ket{00...0}$, $\ket{10..0}$, $\ket{110..0}$ etc., span the low energy eigenspace:
\begin{eqnarray}\label{eqn:Hclock}
&&H_{\clock}=\sum_{t=1}^{T-1}\ketbra{01}{01}_{(t,t+1)}\\
&&=\frac{1}{4}\left[(T-1)\openone+\Z_1-\Z_T-\sum_{t=1}^{T-1}\Z_t\Z_{(t+1)}\right],\nonumber
\end{eqnarray}
where the superscript $(t,t+1)$ indicates the clock qubits acted on by the projection. This Hamiltonian has a simple physical interpretation as a line of ferromagnetically coupled spins with twisted boundary conditions, so that the ground state is spanned by all states with a single domain wall. The term $H_{\text{clockint}}$ applies a penalty $\ket{1}\bra{1}_{t=1}$ to the first qubit to ensure that the clock is in state $\ket{0}^{\otimes T}-$ at time $t=0$.

$H_{\text{prop}}$ acts both on logical and clock qubits. It ensures that the ground state is the history state corresponding to the given circuit. $H_{\text{prop}}$ is a sum of $T$ terms, $H_{\text{prop}} =  \sum_{t=1}^T H_{{\text{prop}},t}$, where each term checks that the propagation from time $t-1$ to $t$ is correct.  For $2 \leq t \leq T-1$, $H_{{\text{prop}},t}$ is defined as:
\begin{eqnarray}\label{eqn:Hprop1}
H_{{\text{prop}},t}&\bydef&\openone \otimes \ketbra{t-1}{t-1} - U_t\otimes\ketbra{t}{t-1}\nonumber\\
&&~- U_t^\dag \otimes \ketbra{t-1}{t} + \openone \otimes\ketbra{t}{t},
\end{eqnarray}
where operators ${\ketbra{t}{t-1}=\ketbra{110}{100}_{(t-1,t,t+1)}}$ etc., act on clock qubits $t-1$, $t$, and $t+1$ and where the operator $U_t$ is the $t^{th}$ gate in the circuit. For the boundary cases ($t=1,T$), one writes $H_{{\text{prop}},t}$ by omitting a clock qubit ($t-1$ and $t+1$ respectively).

We have now explained all the terms in the Hamiltonian from Eq.~\eqref{eqn:htot}---a key building block used to prove the \QMA{}-completeness of $5$-\LH{}~\cite{KSV02}.  The construction reviewed in the present section was also used in a proof of the polynomial equivalence between quantum circuits and adiabatic evolutions~\cite{AvDK+05}.  Which physical systems can implement the Hamiltonian model of computation from Eq.~\eqref{eqn:htot}?  Ideally, we wish to find a simple Hamiltonian that is in principle realizable using current, or near-future technology.  The ground states of many physical systems are real-valued, such as the ground states of the Hamiltonians from Eq.~\eqref{eqn:zzxx} and~\eqref{eqn:zx}.  So a logical first step in our program is to show the \QMA{}-completeness of general real-valued local Hamiltonians.

\subsection{The \QMA-completeness of real-valued Hamiltonians}
Bernstein and Vazirani showed that arbitrary quantum circuits may be represented using real-valued gates operating on real-valued wave functions~\cite{BV97}. Using this idea, one can show that \RFLH{} is already \QMA{}-complete---leaving the proofs in~\cite{KSV02} otherwise intact and changing only the gates used in the circuits.  $H_{\text{in}}$ from Eq.~(\ref{eqn:Hin}) and $H_{\text{clock}}$ from Eq.~(\ref{eqn:Hclock}) are already real-valued and at most $2$-local.  Now consider the terms in $H_{\text{prop}}$ from Eq.~(\ref{eqn:Hprop1}) for the case of {\em self-inverse} elementary gates $U_t=U_t^\dagger$:
\begin{eqnarray}\label{eqn:Hprop}
H_{{\text{prop}},t}&=&\frac{\openone}{4}(\openone-\Z_{(t-1)})(\openone+\Z_{(t+1)})\\
&-&\frac{U}{4}(\openone-\Z_{(t-1)})\X_t(\openone+\Z_{(t+1)})\nonumber
\end{eqnarray}
For the boundary cases ($t=1,T$), define:
\begin{eqnarray}\label{eqn:H1T}
H_{\text{prop,1}} &=& \frac{1}{2}(\openone+\Z_2)-U_1\otimes\frac{1}{2}(\X_1+\X_1\Z_2)\\\nonumber
H_{{\text{prop}},T} &=& \frac{1}{2}(\openone-\Z_{(T-1)})-U_T\otimes\frac{1}{2}(\X_T-\Z_{(T-1)}\X_T).
\end{eqnarray}
The terms from Eq.~(\ref{eqn:Hprop}) and~(\ref{eqn:H1T}) acting on the clock space are already real-valued and at most $3$-local.  As an explicit example of the gates $U_t$, let us define a universal real-valued and self-inverse 2-qubit gate:
\begin{equation*}
R_{ij}(\phi)=\frac{1}{2}(\openone+\PZi)+\frac{1}{2}(\openone-\PZi)\otimes(\sin(\phi)\PXi+ \cos(\phi)\Zj).
\end{equation*}
The gate sequence $R_{ij}(\phi)R_{ij}(\pi/2)$ recovers the universal gate from~\cite{RG02}. This is a continuous set of elementary gates parameterized by the angle $\phi$. Discrete sets of self inverse gates which are universal are also readily constructed. For example, Shi showed that a set comprising the C-NOT plus any one-qubit gate whose square does not preserve the computational basis is universal \cite{SHI02}. We immediately see that a universal set of self-inverse gates cannot contain only the C-NOT and a single one-qubit gate. However, the set $\{$C-NOT$, X,\cos\psi X + \sin\psi Z\}$ is universal for any {\em single} value of $\psi$ which is not a multiple of $\pi/4$.

A reduction from {\sc 5-local} to {\sc 2-local Hamiltonian} was accomplished by the use of \emph{gadgets} that reduced $3$-local Hamiltonian terms to $2$-local terms~\cite{KKR06}.  From the results in~\cite{KKR06} (see also~\cite{OT06}) and the \QMA{}-completeness of \RFLH{}, it now follows that \RH{} is \QMA{}-complete and universal for adiabatic quantum computation. We note that the real product $\PYi\otimes\Yj$, or tensor powers thereof, are not necessary in any part of our construction, and so Hamiltonians composed of the following pairwise products of real-valued Pauli matrices are \QMA{}-complete and universal for adiabatic quantum computation\footnote{The \QMA{}-completeness of this subset of Hamiltonians was found independently by D. Bacon; preprint, (2007).}:
\begin{eqnarray}\label{eqn:resubset}
&&\{{\openone},{\openone}\otimes \X, \openone\otimes\Z,\X\otimes\openone, \\\nonumber
&&~~\Z\otimes\openone,\X\otimes\Z,\Z\otimes\X, \X\otimes\X,\Z\otimes\Z\}.
\end{eqnarray}

To prove our Theorems (1) and (2), we will next show that one can approximate all the terms from Eq.~\eqref{eqn:resubset} using either the ZX or ZZXX Hamiltonians---the Hamiltonians from Eq.~\eqref{eqn:zzxx} and~\eqref{eqn:zx} respectively.  We do this using perturbation theory~\cite{KKR06,OT06} to construct gadget Hamiltonians that approximate the operators $\Z_i\X_j$ and $\PXi\PZi$ with terms from the ZZXX Hamiltonian as well as the operators $\PZi\PZi$ and $\PXi\Xj$ with terms from the ZX Hamiltonian.

\subsection{The ZZXX gadget} We use the ZZXX Hamiltonian from Eq.~(\ref{eqn:zzxx}) to construct the interaction $\PZi\Xj$ from $\sigma^x\sigma^x$ and $\sigma^z\sigma^z$ interactions. Let $H_\text{eff} =  \alpha_{ij}\PZi\Xj\otimes\ket{0}\bra{0}_k$, where qubit $k$ is an ancillary qubit and define the penalty Hamiltonian $H_{\text{p}}$ and corresponding Green's function $G(z)$ as follows:
\begin{eqnarray}\label{penalty}
H_{\text{p}} &=& \delta\ket{1}\bra{1}_k=\frac{\delta}{2}(\openone-\Zk)\qquad  \text{and}\\
G(z)&\bydef&(z\openone- H_{\text{p}})^{-1}\nonumber.
\end{eqnarray}

$H_{\text{p}}$ splits the Hilbert space into a degenerate low energy eigenspace ${\cal L}_-=\text{span}\{\ket{s_is_j}\ket{0}|\forall s_i,s_j\in\{0,1\}\}$, in which qubit $k$ is $\ket{0}$, and a $\delta$ energy eigenspace ${\cal L}_+=\text{span}\{\ket{s_is_j}\ket{1}|\forall s_i,s_j\in\{0,1\}\}$, in which qubit $k$ is $\ket{1}$.

\begin{figure}[h]
\center
\includegraphics[width=6cm]{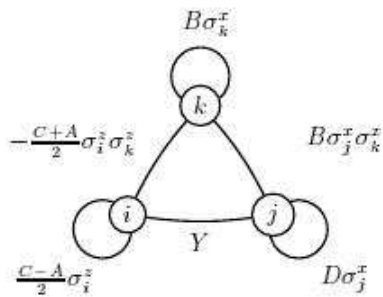}
\caption{The ZZXX gadget used to approximate the operator $\PZi\Xj$ using only $\sigma^x\sigma^x$ and $\sigma^z\sigma^z$ interactions.  The present figure presents a diagrammatic representation of the Perturbation Hamiltonian $V=V_1+V_2+V_3$ from Eq.~\eqref{eqn:V} applied to qubits $i,j$ and $k$. Not shown in the present figure is an overall constant energy shift of $D$.}\label{fig:one}
\end{figure}

First, we give the ZZXX Hamiltonian which produces an effective $\sigma^z\sigma^z$ interaction in the low energy subspace. Let $Y$ be an arbitrary ZZXX Hamiltonian acting on qubits $i$ and $j$ and consider a perturbation $V=V_1+V_2+V_3$ that breaks the ${\cal L}_-$ zero eigenspace degeneracy by creating an operator $\mathcal{O}(\epsilon)$ close to $H_\text{eff}$ in this space:
\begin{eqnarray}\label{eqn:V}
V_1 &=& [Y + D(\Xj+\openone)]\otimes\openone_k - A\PZi\otimes\ket{0}\bra{0}_k\nonumber\\
V_2 &=& B(\Xj + \openone )\otimes\Xk\\
V_3 &=& C\PZi\otimes\ket{1}\bra{1}_k\nonumber.
\end{eqnarray}
The term $V_2$ above allows the mediator qubit $k$ to undergo \emph{virtual excitations} and applies an $\X$ term to qubit $j$ during transitions between the ${\cal L}_-$ and ${\cal L}_+$ subspaces.  During excitation into ${\cal L}_+$, the term $V_3$ applies a $\Z$ term to qubit $i$. This perturbation is illustrated in figure~\ref{fig:one}

Let $\Pi_{\pm}$ be projectors on ${\cal L}_{\pm}$; for arbitrary operator $O$ we define $O_{\pm \mp}=\Pi_{\pm} O \Pi_{\mp}$ ($O_{\pm \pm}=\Pi_{\pm} O \Pi_{\pm}$) and let $\lambda(O)$ denote the lowest eigenvalue of $O$.  One approximates $\lambda(H_{\text{targ}})$ of the desired low energy effective 2-local Hamiltonian by a realizable 2-local physical Hamiltonian $\tilde{H}=H_p+V$, where $\lambda(\tilde{H})$ is calculated using perturbation theory.  The spectrum of $\tilde{H}_{--}$ is approximated by the projection of the self-energy operator $\Sigma(z)$ for real-valued $z$ which has the following series expansion:
\begin{eqnarray}\label{eqn:selfenergy}
\Sigma_{--}(z) &=& \overbrace{H_{\text{p--}}}^{\text{$0^{th}$}} + \overbrace{V_{--}}^{\text{$1^{st}$}} +\overbrace{V_{-+} G_{++}(z)V_{+-}}^{\text{$2^{nd}$}}\\\nonumber
&+& \underbrace{V_{-+} G_{++}(z) V_+ G_{++}(z) V_{+-}}_{\text{$3^{rd}$}}\\\nonumber
&+&\mathcal{O}\left(\|V\|^4\delta^{-3}\right)+ \cdots\nonumber
\end{eqnarray}

Note that with our penalty Hamiltonian $H_{--}=0$, and for the perturbing Hamiltonian $V = V_1 + V_2 + V_3$ only $V_1$ is nonzero in the low energy subspace, $V_1$ and $V_3$ are nonzero in the high energy subspace, and only $V_2$ induces transitions between the two subspaces. The non-zero projections are:
\begin{eqnarray}
{V_1}_{--} &=& [Y + A \PZi + D(\Xj + \openone )]\otimes \ket{0}\bra{0}_k\nonumber\\
{V_2}_{-+} &=& B(\Xj + \openone )\otimes\ket{0}\bra{1}_k\nonumber\\
{V_2}_{+-} &=& B(\Xj + \openone )\otimes\ket{1}\bra{0}_k\\
{V_3}_{++} &=& V_3\nonumber\\
V_{+\phantom{3+}}&=&(Y+C\PZi+D(\Xj + \openone ))\otimes \ket{1}\bra{1}_k\nonumber
\end{eqnarray}
The series expansion of the self-energy follows directly:
\begin{eqnarray}\nonumber
1^{st}&:& (Y - A \PZi+D(\Xj + \openone ))\otimes \ket{0}\bra{0}_k\nonumber\\
2^{nd}&:&\frac{B^2}{z-\delta}(\Xj+\openone)^2\otimes\ket{0}\bra{0}_k\\
3^{rd}&:& \frac{B^2C}{(z-\delta)^2}(\Xj+\openone)\PZi (\Xj+\openone)\otimes\ket{0}\bra{0}_k\nonumber\\
\phantom{3^{rd}}&+&\frac{B^2}{(z-\delta)^2}(\Xj+\openone)Y(\Xj+\openone)\otimes\ket{0}\bra{0}_k\nonumber\\
\phantom{3^{rd}}&+&\frac{4DB^2}{(z-\delta)^2}(\Xj+\openone)^3\otimes\ket{0}\bra{0}_k\nonumber
\end{eqnarray}

The self-energy in the low energy subspace (where qubit $k$ is in state $\ket{0}$) is therefore:
\begin{eqnarray}\label{SEZZXX1}
\Sigma_{--}(z)&\simeq&\tilde{Y} +\left(\frac{2B^2C}{(z-\delta)^2}- A\right) \PZi\\\nonumber
&+&\left(\frac{2B^2}{z-\delta} + D + \frac{4DB^2}{(z-\delta)^2}\right)(\Xj+\openone)\\\nonumber
&+& \frac{2B^2C}{(z-\delta)^2}\PZi\Xj\\\nonumber
&+&\mathcal{O}\left(\|V\|^4\delta^{-3}\right)+ \cdots\nonumber
\end{eqnarray}
$\tilde{Y}$ is the interaction between qubits $i$ and $j$ which is the original physical interaction dressed by the effect of virtual excitations into the high energy subspace.
\begin{equation}
\tilde{Y} = Y  + \frac{B^2}{(z-\delta)^2}(\Xj+\openone)Y(\Xj+\openone)
\end{equation}

In practice there will always be some interaction between qubits $i$ and $j$. We assume $Y$ is a ZZXX Hamiltonian and express the dressed Hamiltonian $\tilde{Y}$ in terms of modified coupling coefficients. Writing the physical Hamiltonian:
\begin{eqnarray}
Y &=& h_i\PZi+h_j\Zj+\Delta_i\PXi+\Delta_j\Xj+\\\nonumber
&+&J_{ij}\PZi\Zj+K_{ij}\PXi\Xj.
\end{eqnarray}
The new dressed coupling strengths are:
\begin{eqnarray}
h_i &\mapsto& h_i\left(1 + \frac{2B^2}{(z-\delta)^2}\right) \\\nonumber
\Delta_i &\mapsto& \Delta_i \left(1 + \frac{2B^2}{(z-\delta)^2}\right) +\frac{2B^2}{(z-\delta)^2}K_{ij}\\\nonumber
\Delta_j &\mapsto& \Delta_j \left(1 + \frac{2B^2}{(z-\delta)^2}\right)\\\nonumber
K_{ij} &\mapsto& K_{ij} \left(1 + \frac{2B^2}{(z-\delta)^2}\right)  + \frac{2B^2}{(z-\delta)^2}\Delta_i
\end{eqnarray}
with additional couplings:
\begin{eqnarray}
\frac{2B^2}{(z-\delta)^2}\Delta_j \openone + \frac{2B^2}{(z-\delta)^2}h_i\PZi\Xj
\end{eqnarray}
We see that the effect of the gadget on any existing physical interaction is to modify the coupling constants, add an overall shift in energy, and to add a small correction to the $\sigma^z\sigma^z$ coupling which depends on the strength of the $\PZi$ term in $Y$. If $Y$ is regarded as the net uncontrolled physical Hamiltonian coupling $i$ and $j$ (a source of error) it is only the local $\PZi$ field which contributes to an error in the $\sigma^z\sigma^z$ coupling strength.

We make the following choices for our gadget parameters $A$, $B$, $C$ and $D$:
\begin{eqnarray}
A &=& \alpha_{ij}\\\nonumber
B &=& \left(\frac{\delta}{\bar E}\right)^{2/3} \bar E\\\nonumber
C &=& \frac{\alpha_{ij}}{2}\left(\frac{\delta}{\bar E}\right)^{2/3} \\\nonumber
D &=& 2\delta^{1/3}{\bar E}^{2/3} \nonumber
\end{eqnarray}
Where $\bar E$ is an energy scale parameter to be fixed later.

We expand the self-energy~(\ref{SEZZXX1}) in the limit where $z$ is constant ($z=\mathcal{O}(1)\ll\delta$). Writing $(z-\delta)^{-1}\simeq-\frac{1}{\delta}+\mathcal{O}(\frac{1}{\delta^2})$ gives:
\begin{eqnarray}\label{SEZZXX2}
\Sigma(0)_{--} &=& \tilde Y + \alpha_{ij} \PZi\Xj\\
&+& 8\frac{\bar E^{4/3}}{\delta^{1/3}}(\Xj+\openone)\nonumber\\
&+&\mathcal{O}\left(\|V\|^4\delta^{-3}\right)+ \cdots\nonumber
\end{eqnarray}

For the self-energy to become $\mathcal{O}(\epsilon)$ close to $Y + \alpha_{ij}\PZi\Xj\otimes\ket{0}\bra{0}_k$, the error terms in~(\ref{SEZZXX2}) must be bounded above by $\epsilon$ through an appropriate choice of $\delta$. Define a lower bound on the spectral gap $\delta$ as an inverse polynomial in $\epsilon$: $\delta\geq \bar E\epsilon^{-r}$, where $\bar E$ is a constant and integer $r\geq1$.  Now bound $r$ by considering the (weak) upper bound on  $\|V\|$:
\begin{eqnarray}
\|V\|&\leq&\| Y\| + |\alpha_{ij}| + 4\delta^{1/3}\bar E^{2/3}\\\nonumber
&+& 2\bar E\left(\frac{\delta}{\bar E}\right)^{2/3} + \frac{|\alpha_{ij}|}{2}\left(\frac{\delta}{\bar E}\right)^{2/3}.\\\nonumber
\end{eqnarray}
The largest term in $\delta^{-3}\|V\|^4$ is  $\mathcal{O}(\bar E (\bar E/\delta)^{1/3})$, and so in order that $\delta^{-3}\|V\|^4<\epsilon$ we require $r\geq3$. This also bounds the term below fourth order, $\bar E^{4/3}\delta^{-1/3}=\mathcal{O}(\bar E \epsilon)$  and so for $z\ll\delta$ we obtain $\|\Sigma_{--}(z)-H_{\text{eff}}\| = \mathcal{O}(\epsilon)$. In fact, $\Sigma(0)_{--} = H_{\text{eff}} + \bar E \epsilon (\Xj+\openone)$. Now apply Theorem (3) from~\cite{KKR06} and it follows that $|\lambda(H_{\text{eff}})-\lambda(\tilde{H})|=\mathcal{O}(\epsilon)$. It also follows from Lemma (11) of~\cite{KKR06} that the ground state wavefunction of $H_{\text{eff}}$ is also close to the ground state of our gadget.

The ZZXX Hamiltonian~\eqref{eqn:zzxx} allows for the direct realization of all terms in \eqref{eqn:resubset} except for $\Z\X$ and $\X\Z$ interactions.  These terms can be approximated with only $\mathcal{O}(\epsilon)$ error using the gadget in the present section---thereby showing that the ZZXX Hamiltonian can efficiently approximate all terms from~\eqref{eqn:resubset}.  Similarly, the ZX Hamiltonian allows for the direct realization of all terms in \eqref{eqn:resubset} except for $\Z\Z$ and $\X\X$ interactions.  These terms will be approximated with only $\mathcal{O}(\epsilon)$ error by defining gadgets in the coming sections---showing that the ZX Hamiltonian can also be used to efficiently approximate all terms from~\eqref{eqn:resubset}.

\subsection{The ZZ from ZX gadget} We approximate the operator $\beta_{ij}\PZi\Zj$ using the ZX Hamiltonian in Eq.~(\ref{eqn:zx}) by defining a penalty Hamiltonian as in Eq.~(\ref{penalty}). The required perturbation is a sum of terms $V = V_1 + V_2$:
\begin{eqnarray}\label{eqn:zzfromzx}
V_1&=& Y + A\ket{0}\bra{0}_k\\
V_2 &=& B(\PZi - \Zj)\otimes\Xk\nonumber
\end{eqnarray}

\begin{figure}[h]
\center
\includegraphics[width=6cm]{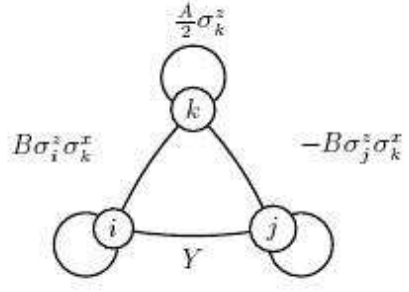}
\caption{The $ZZ$ from $ZX$ gadget: The present figure presents a diagrammatic representation of the Perturbative Hamiltonian $V = V_1 + V_2$ from Eq.~\eqref{eqn:zzfromzx} applied to qubits $i,j$ and $k$. In addition to these terms shown in the present figure, there is an overall energy shift of $A/2$.}\label{fig:two}
\end{figure}

The non-zero projections are:
\begin{eqnarray}
V_{1++}&=& Y\otimes\ket{1}\bra{1}_k\\
V_{1--} &=& (Y +A\openone\otimes\ket{0}\bra{0}_k\nonumber\\
V_{2+-}&=&B(\PZi-\Zj)\otimes\ket{1}\bra{0}_k\nonumber\\
V_{2-+} &=&B(\PZi-\Zj)\otimes\ket{0}\bra{1}_k\nonumber
\end{eqnarray}
$V_1$ does not couple the low and high energy subspaces and $V_2$ couples the subspaces but is zero in each subspace.
The series expansion of the self-energy follows directly:
\begin{eqnarray}\nonumber
1^{st}&:& (Y +A\openone)\otimes\ket{0}\bra{0}_k\\
2^{nd}&:&\frac{B^2(\PZi-\Zj)^2}{(z-\delta)}\otimes\ket{0}\bra{0}_k\\\nonumber
3^{rd}&:& \frac{B^2}{(z-\delta)^2}(\PZi-\Zj)Y(\PZi-\Zj) \otimes\ket{0}\bra{0}_k\nonumber
\end{eqnarray}
Note that in this case the desired terms appear at second order in the expansion, rather than at third order as was the case for the ZX from ZZXX gadget. The terms which dress the physical hamiltonian $Y$ coupling qubits $i$ and $j$ appear at third order. The series expansion of the self-energy in the low energy subspace is:
\begin{equation}
\begin{split}
\Sigma(z)_{--} &= (\tilde Y  +A\openone)\\
&+\frac{2B^2(1-\PZi\Zj)}{(z-\delta)}\\
&+{\cal O}(||V||^4\delta^{-3})\\
\end{split}
\end{equation}
where the dressed interaction $\tilde{Y}$ is defined:
\begin{equation}
\tilde Y = Y +  \frac{B^2}{(z-\delta)^2}(\PZi-\Zj)Y(\PZi-\Zj)
\end{equation}
We assume that the physical interaction $Y$ between $i$ and $j$ qubits is a ZX Hamiltonian and express the dressed Hamiltonian in terms of modified coupling constants. Writing the physical Hamiltonian:
\begin{eqnarray}\label{ZXHAM}
Y &= h_i\PZi + h_j\Zj + \Delta_i\PXi+\Delta_j\Xj\\
&+ J_{ij}\PZi\Xj +K_{ij}\PXi\Zj\nonumber
\end{eqnarray}
We obtain modified coupling strengths:
\begin{eqnarray}
h_i&\mapsto&h_i + \frac{2B^2(h_i-h_j)}{(z-\delta)^2}\\
h_j&\mapsto&h_i + \frac{2B^2(h_j-h_i)}{(z-\delta)^2}.\nonumber
\end{eqnarray}
In this case only the local Z field strengths are modified.

We choose values for the perturbation interaction strengths as follows: $B = \sqrt{\frac{\beta_{ij}\delta}{2}}$ and $A = \beta_{ij}$ and expand the self-energy in the limit where $z$ is constant ($z=\mathcal{O}(1)\ll\delta$):
\begin{equation}
\begin{split}
\Sigma(0)_{--} &= \tilde Y + \beta_{ij}\PZi\Zj\\
& +{\cal O}(||V||^4\delta^{-3}).\\
\end{split}
\end{equation}
We again choose $\delta$ to be an inverse power in a small parameter $\epsilon$ so that $\delta \geq \bar E \epsilon^{-s}$, and again use the (weak) upper bound on $||V||$:
\begin{equation}
||V|| \leq ||Y|| +\beta_{ij} +\sqrt{2\beta_{ij}\delta}
\end{equation}
The largest term in $||V||^4\delta^{-3}$ is $4\beta_{ij}^{2}\delta^{-1}$, and so in order that $||V||^4\delta^{-3}< \epsilon$ we require $r\geq1$.

Using the gadget defined in the present section, the ZX Hamiltonian can now be used to efficiently approximate all terms in \eqref{eqn:resubset} except for $\X\X$ interactions.  These interactions can also be approximated with only $\mathcal{O}(\epsilon)$ error by defining an additional gadget in the next section.


\subsection{The XX from ZX gadget} An $\sigma^x\sigma^x$ coupling may be produced from the $\sigma^z\sigma^x$ coupling as follows. We define a penalty Hamiltonian and corresponding Green's function:
\begin{eqnarray}
H_p &= &\frac{\delta}{2}(\openone-\Xk)=\delta\ket{-}\bra{-}\nonumber\\
G_{++}&=&\frac{1}{z-\delta}\ket{-}\bra{-}_k.
\end{eqnarray}
This penalty Hamiltonian splits the Hilbert space into a low energy subspace in which the ancilla qubit $k$ is in state $\ket{+} = (\ket{0}+\ket{1})/\sqrt{2}$ and a high energy subspace in which the ancilla qubit $k$ is in state $\ket{-} = (\ket{0}-\ket{1})/\sqrt{2}$.

\begin{figure}[h]
\center
\includegraphics[width=6cm]{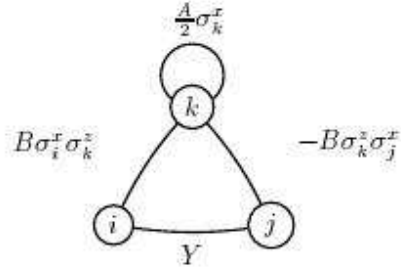}
\caption{The $XX$ from $ZX$ gadget: The present figure presents a diagrammatic representation of the Perturbative Hamiltonian $V=V_1+V_2$ from Eq.~\eqref{eqn:xxfromzx} applied to qubits $i,j$ and $k$. In addition to the terms shown in the present figure, there is an overall energy shift of $A/2$. The penalty term applied to qubit $k$ is the $\X$ basis.}\label{fig:three}
\end{figure}

The perturbation is a sum of two terms $V=V_1+V_2$, where $V_1$ and $V_2$ are given by:
\begin{eqnarray}\label{eqn:xxfromzx}
V_1 &=& Y\otimes\openone_k + A\ket{+}\bra{+}_k\\
V_2 &=& B(\PXi-\Xj)\Zk\nonumber
\end{eqnarray}
The non-zero projections are:
\begin{eqnarray}
V_{1++} &=& \ket{-}\bra{-}V_1\ket{-}\bra{-}_k\\
&=& Y\otimes\ket{-}\bra{-}_k\nonumber\\
V_{1--}&=& Y\otimes\ket{+}\bra{+}_k + A \ket{+}\bra{+}_k\nonumber\\
V_{2+-}&=&\ket{-}\bra{-}V_2\ket{+}\bra{+}_k\nonumber\\
&=& B(\PXi-\Xj)\ket{-}\bra{+}_k\nonumber\\
V_{2-+}&=&\ket{+}\bra{+}V_2\ket{-}\bra{-}_k\nonumber\\
&=& B(\PXi-\Xj)\ket{+}\bra{-}_k.\nonumber
\end{eqnarray}
Once more we see that the perturbation $V_1$ does not couple the subspaces, whereas $V_2$ couples the subspaces but is zero in each subspace. This perturbation is illustrated in Figure~\ref{fig:three}.

The series expansion of the self-energy follows:
\begin{eqnarray}
1^{st}&:& (Y +A\openone)\otimes\ket{+}\bra{+}_k\\
2^{nd}&:&\frac{B^2(\PXi-\Xj)^2}{(z-\delta)}\otimes\ket{+}\bra{+}_k\nonumber\\
3^{rd}&:& \frac{B^2}{(z-\delta)^2}(\PXi-\Xj)Y(\PXi-\Xj)\nonumber
\end{eqnarray}
Again we see that the desired term appears at second order, while the third order term is due to the dressing of the physical interaction $Y$ between qubits $i$ and $j$. In the low energy subspace the series expansion of the self-energy to third order is:
\begin{eqnarray}
\Sigma(z)_{--}&=& (\tilde Y +A\openone)\\\nonumber
&+&\frac{2B^2(\openone-\PXi\Xj)}{(z-\delta)}\\\nonumber
&+&{\cal O}(||V||^4\delta^{-3})\nonumber
\end{eqnarray}
where the dressed interaction $\tilde{Y}$ is defined:
\begin{equation}
\tilde Y = Y +  \frac{B^2}{(z-\delta)^2}(\PXi-\Xj)Y(\PXi-\Xj).
\end{equation}
Once more we assume the physical Hamiltonian $Y$ is a ZX Hamiltonian~\ref{ZXHAM} and we describe the effects of dressing to low order in terms of the new dressed coupling strengths:
\begin{eqnarray}
\Delta_i&\mapsto&\Delta_i + \frac{2B^2(\Delta_i-\Delta_j)}{(z-\delta)^2}\\
\Delta_j&\mapsto&\Delta_i + \frac{2B^2(\Delta_j-\Delta_i)}{(z-\delta)^2}\nonumber,
\end{eqnarray}
and in this case only the local X field strengths are modified.

Choosing values for our gadget parameters $A = \gamma_{ij}$ and $B = \sqrt{\frac{\gamma_{ij}\delta}{2}}$ and expanding the self-energy in the limit where $z$ is constant ($z=\mathcal{O}(1)\ll\delta$) gives:
\begin{eqnarray}
\Sigma(0)_{--}&=& \tilde Y\otimes\ket{+}\bra{+}_k\\\nonumber
&+&\gamma_{ij}\PXi\Xj\otimes\ket{+}\bra{+}_k\\\nonumber
&+& {\cal O}(||V||^4\delta^{-3})\nonumber
\end{eqnarray}
As before, this self-energy may be made $\mathcal{O}(\epsilon)$ close to the target Hamiltonian by a bound $\delta\geq \bar E\epsilon^{-1}$.

\paragraph{Summary} The proof of Theorem (1) follows from the simultaneous application of the ZZXX gadget illustrated in Fig.~\ref{fig:one} to realize all $\Z\X$ terms in the target Hamiltonian using a ZZXX Hamiltonian. Similarly, application of the two gadgets illustrated in Fig.~\ref{fig:two} and~\ref{fig:three} to realize $\X\X$ and $\Z\Z$ terms in the target Hamiltonian proves the first part of Theorem (2).  Our result is based on Theorem (3) from~\cite{KKR06} which allowed us to approximate (with $\mathcal{O}(\epsilon)$ error) all the Hamiltonian terms from Eq.~\eqref{eqn:htot} using either the ZZXX or ZX Hamiltonians.  It also follows from Lemma (11) of~\cite{KKR06} that the ground state wavefunction of $H_{\text{eff}}$ is also close to the ground state of our gadget.  So to complete our proof, it is enough to show that each gadget satisfies the criteria given in Theorem (3) from~\cite{KKR06}.

\section{Conclusion}
The objective of this work was to provide simple model Hamiltonians that are of practical interest for experimentalists working towards the realization of a universal adiabatic quantum computer.  Accomplishing such as task also enabled us to find the simplest known \QMA-complete 2-local Hamiltonians.  The $\X\X$ coupler is realizable using systems including capacitive coupling of flux qubits~\cite{AB03} and spin models implemented with polar molecules~\cite{MBZ06}. In addition, a $\sigma_z\sigma_x$ coupler for flux qubits is given in~\cite{OMT+99}. The ZX and ZZXX Hamiltonians enable gate model~\cite{BDEJ95}, autonomous~\cite{J05}, measurement-based~\cite{BR06} and universal adiabatic quantum computation~\cite{AvDK+05,KKR06,OT06}, and may also be useful for quantum annealing~\cite{SNS07}.  For these reasons, the reported Hamiltonians are of interest to those concerned with the practical construction of a universal adiabatic quantum computer\footnote{We thank C.J.S. Truncik, R.G. Harris, W.G. Macready, M.H.S. Amin, A.J. Berkley, P. Bunyk, J. Lamothe, T. Mahon and G. Rose.  J.D.B. and P.J.L. completed parts of this work well on staff as D-Wave Systems Inc.}.


\end{document}